\documentclass[12pt]{iopart}

\usepackage{amssymb}
\usepackage{bm}
\usepackage{bbold}

\begin{document}

\title[Casimir invariants and the Jacobi identity in Dirac's theory]{Casimir invariants and the Jacobi identity in Dirac's theory of constrained Hamiltonian systems}

\author{C. Chandre}
\address{Centre de Physique Th\'eorique, CNRS -- Aix-Marseille Universit\'e, Campus de Luminy, case 907, 13009 Marseille, France}
\ead{chandre@cpt.univ-mrs.fr}

\begin{abstract}
We consider constrained Hamiltonian systems in the framework of Dirac's theory. We show that the Jacobi identity results from imposing that the constraints are Casimir invariants, regardless of the fact that the matrix of Poisson brackets between constraints is invertible or not. We point out that the proof we provide ensures the validity of the Jacobi identity everywhere in phase space, and not just on the surface defined by the constraints. Two examples are considered: A finite dimensional system with an odd number of constraints, and the Vlasov-Poisson reduction from Vlasov-Maxwell equations. 
\end{abstract}

\maketitle



\section{Introduction}

Imposing constraints on a Hamiltonian system is routinely done using Dirac's theory~\cite{dira50,dira58,Bhans76,Bsund82}. The key point is to compute the matrix ${\mathbb C}$ of the Poisson brackets between two constraints (i.e., whose elements are $C_{nm}=\{\Phi_n,\Phi_m\}$ where $\Phi_n({\bf z})=0$ are the constraints) and to invert this matrix. Under this hypothesis, it has been shown that the following bracket, called Dirac bracket,
\begin{equation}
\label{eq:DB}
\{F,G\}_*=\{F,G\}-\{F,\Phi_n\}D_{nm}\{\Phi_m,G\},
\end{equation}
where ${\mathbb D}={\mathbb C}^{-1}$, is a Poisson bracket. The technical difficulty is to prove the Jacobi identity, and this has been done by Dirac in Ref.~\cite{dira50}, and his proof relies heavily on the invertibility of ${\mathbb C}$ (see Eq.~(59) which results from Eq.~(35) in Ref.~\cite{dira50}, or Appendix B in Ref.~\cite{morr09}). As a consequence of the fact that ${\mathbb D}$ is the inverse of ${\mathbb C}$, the constraints are Casimir invariants, i.e., $\{\Phi_n,G\}_*=0$ for any observable $G$. 

What if the matrix ${\mathbb C}$ is not invertible? This could happen for instance if there is an odd number of constraints (since ${\mathbb C}$ is antisymmetric) or if the constraints are reducible or redundant~\cite{Bhenn92,deri96,abre02,bizd07}. The purpose of the present article is to address this question. Our main result is to show that the Jacobi identity is a property which results from imposing that the constraints are Casimir invariants, regardless of the invertibility of ${\mathbb C}$. This result holds for canonical or non-canonical Hamiltonian systems (see Refs.~\cite{Bmars02,morr82,morr98} for an introduction to non-canonical Hamiltonian systems). We notice that the proof we provide shows the validity of the Jacobi identity everywhere in phase space, and not just on the surface defined by the constraints as found in Refs.~\cite{Bhenn92,abre02,bizd07}.    
   
We consider a finite dimensional Hamiltonian system whose variables are denoted ${\bf z}=(z_1,z_2,\ldots, z_N)$. It is given by a Hamiltonian $H({\bf z})$, a scalar function of the variables, and a Poisson bracket written as
\begin{equation}
\label{eq:PB}
\{F,G\}=\frac{\partial F}{\partial {\bf z}}\cdot {\mathbb J}({\bf z})\frac{\partial G}{\partial {\bf z}},
\end{equation}
where the Poisson matrix ${\mathbb J}({\bf z})$ is such that the bracket~(\ref{eq:PB}) is antisymmetric  and satisfies the Jacobi identity [in addition to the bilinearity and the Leibnitz rule which are already ensured by the form of the bracket~(\ref{eq:PB})]. 

We impose a set of $M<N-2$ constraints $\Phi_m({\bf z})=0$ for $m=1,\ldots,M$ which are scalar functions of the variables ${\bf z}$. We consider brackets of the form~(\ref{eq:DB}) with an antisymmetric matrix ${\mathbb D}$ which is not necessarily the inverse of ${\mathbb C}$ (and consequently these brackets are not not Poisson brackets in general).
The matrix associated with the bracket~(\ref{eq:DB}) is given by~\cite{chan13}
\begin{equation}
{\mathbb J}_*={\mathbb J}-{\mathbb J}\hat{\cal Q}^\dagger{\mathbb D} \hat{\cal Q} {\mathbb J},
\end{equation}
where the matrix $\hat{\cal Q}$ has elements $\hat{\cal Q}_{ni}=\partial \Phi_n/\partial z_i$. The matrix ${\mathbb D}$ is chosen such that the constraints are Casimir invariants. This leads to the following condition on ${\mathbb D}$~: 
\begin{equation}
{\mathbb J} \hat{\cal Q}^\dagger ({\mathbb 1}-{\mathbb D} {\mathbb C})=0,
\label{eq:forD}
\end{equation}
where ${\mathbb 1}$ is the $M\times M$ identity matrix and ${\mathbb C}=\hat{\cal Q}{\mathbb J}\hat{\cal Q}^\dagger$.
A first situation is when ${\mathbb C}$ is invertible, and hence a possible solution to Eq.~(\ref{eq:forD}) is ${\mathbb D}={\mathbb C}^{-1}$. This is the main case considered in the literature, for which the Jacobi identity is proved in order to ensure that the Dirac bracket is a Poisson bracket~\cite{dira50,morr09}. 

However in many instances, the matrix ${\mathbb C}$ is not invertible, and, to the best of our knowledge, there is no proof of the Jacobi identity in these cases which holds everywhere in phase space. There are two properties which are strongly dependent on the particular choice of matrix ${\mathbb D}$, which are $(i)$ the Jacobi identity and $(ii)$ the fact that the constraints are Casimir invariants. The purpose of this article is to show that $(ii)$ implies $(i)$, or in other words that imposing that the constraints are Casimir invariants, i.e., Eq.~(\ref{eq:forD}), is sufficient to ensure that the Dirac bracket~(\ref{eq:DB}) is a Poisson bracket. The proof is done in Sec.~\ref{sec:JI}. In Sec.~\ref{sec:examples} we illustrate the computation of the Dirac bracket in cases where ${\mathbb C}$ is not invertible, with two examples: a finite-dimensional Hamiltonian system with an odd number of constraints, and the Vlasov-Poisson reduction from Vlasov-Maxwell equations. 

Before going into the proof of the Jacobi identity, we would like to address two questions: 

1) Does the Jacobi identity for the bracket~(\ref{eq:DB}) imply that the constraints are Casimir invariants, i.e., does $(i)$ implies $(ii)$? The answer is no, and we provide a counter example below. Consider a Poisson matrix ${\mathbb J}$ written in block form as
$$
{\mathbb J}=\left(\begin{array}{cc} {\mathbb C} & 0 \\ 0 & \bar{\mathbb J}\end{array}\right),
$$
where ${\mathbb C}$ and $\bar{\mathbb J}$ satisfy individually the Jacobi identity so that the matrix ${\mathbb J}$ too. The bracket~(\ref{eq:DB}) is characterized by a matrix 
$$
{\mathbb J}_*=\left(\begin{array}{cc} {\mathbb C}({\mathbb 1}-{\mathbb D}{\mathbb C}) & 0 \\ 0 & \bar{\mathbb J}\end{array}\right),
$$
for $\hat{\cal Q}=({\mathbb 1},\, 0)$. Assuming that ${\mathbb C}$ is invertible, we choose ${\mathbb D}={\mathbb C}^{-1}(1-\lambda)$ so that $\tilde{\mathbb C}={\mathbb C}({\mathbb 1}-{\mathbb D}{\mathbb C})=\lambda {\mathbb C}$ which satisfies the Jacobi identity inherited from ${\mathbb C}$. In order to have the constraints as Casimir invariants, $\tilde{\mathbb C}$ must vanish. Therefore the bracket~(\ref{eq:DB}) satisfies the Jacobi identity in this case, but does not have the constraints $(z_1,\ldots,z_M)$ (where $M$ is the number of columns of ${\mathbb C}$) as Casimir invariants.  

2) Given that Eq.~(\ref{eq:forD}) might have more than one solution, does it lead to different expressions for the Dirac bracket? The answer is no. We consider ${\mathbb D}$ a solution of Eq.~(\ref{eq:forD}). We notice that any matrix $\tilde{\mathbb D}={\mathbb D}+{\mathbb \Delta}$ with ${\mathbb J}\hat{\cal Q}^\dagger {\mathbb \Delta} {\mathbb C}=0$ also satisfies Eq.~(\ref{eq:forD}). The Dirac bracket is obtained from the Dirac projector~\cite{chan13} ${\cal P}=1-{\mathbb J}\hat{\cal Q}^\dagger {\mathbb D} \hat{\cal Q}$, i.e., ${\mathbb J}_*={\cal P}{\mathbb J}{\cal P}^\dagger$. If we consider the other projector associated with $\tilde{\mathbb D}$, i.e., $\tilde{\cal P}={\cal P}-{\mathbb J}\hat{\cal Q}^\dagger {\mathbb \Delta}\hat{\cal Q}$, then we show that 
$$
\tilde{\cal P}{\mathbb J}\tilde{\cal P}^\dagger={\cal P}{\mathbb J}{\cal P}^\dagger,
$$
where we use the identity ${\cal P}{\mathbb J}\hat{\cal Q}^\dagger=0$. This identity ensures that the Dirac bracket is unique, even if there might be more than one solution to Eq.~(\ref{eq:forD}).

\section{Proof of Jacobi identity}
\label{sec:JI}

A proof of a weak version of the Jacobi identity, i.e., the validity of the Jacobi identity on the surface defined by the constraints, has been detailed, e.g., in Refs.~\cite{Bhenn92,abre02,bizd07}. It is based on showing that
$$
\{F,\{G,H\}_*\}_*\approx \{F',\{G',H'\}\},
$$ 
where $F'=F-\{F,\Phi_n\}D_{nm}\Phi_m$ in order to deduce that
$$
\{F,\{G,H\}_*\}_*+\{H,\{F,G\}_*\}_*+\{G,\{H,F\}_*\}_*\approx 0.
$$
Here we show that the Jacobi identity holds everywhere in phase space, i.e., the weak equality can be made a strong one, i.e.,
$$
\{F,\{G,H\}_*\}_*+\{H,\{F,G\}_*\}_*+\{G,\{H,F\}_*\}_*= 0.
$$
As a consequence, even in the case where the matrix ${\mathbb C}$ is not invertible, the Dirac bracket~(\ref{eq:DB}) [if it can be constructed using Eq.~(\ref{eq:forD})] is a Poisson bracket. 

First we perform a local change of coordinates such that the new variables are the constraint functions. This can be done at least locally under the assumption of the change of coordinates. In other terms, we assume that $\Phi_k({\bf z})=z_k$ for $k\in[1,M]$. 
We assume that ${\mathbb J}$ satisfies the Jacobi identity. In the variables $({\bm \Phi},{\bm w})$ the Poisson matrix is expressed by blocks
$$
{\mathbb J}=\left( \begin{array}{cc} 
{\mathbb C} & -{\mathbb B}^\dagger\\
{\mathbb B} & \bar{\mathbb J}
\end{array}\right).
$$
The Poisson matrix associated with the bracket~(\ref{eq:DB}) is given by
$$
{\mathbb J}_*=\left( \begin{array}{cc} 
{\mathbb C}({\mathbb 1}-{\mathbb D}{\mathbb C}) & -({\mathbb 1}-{\mathbb C}{\mathbb D}){\mathbb B}^\dagger\\
{\mathbb B}({\mathbb 1}-{\mathbb D}{\mathbb C}) & \bar{\mathbb J}+{\mathbb B}{\mathbb D}{\mathbb B}^\dagger
\end{array}\right).
$$
We assume that ${\mathbb D}$ is chosen such that the constraints are Casimir invariants of ${\mathbb J}_*$. This condition becomes:
\begin{eqnarray}
&& {\mathbb C}({\mathbb 1}-{\mathbb D}{\mathbb C})=0,\label{eq:Cas1}\\
&& {\mathbb B}({\mathbb 1}-{\mathbb D}{\mathbb C})=0.\label{eq:Cas2}
\end{eqnarray}

The Jacobi identity for ${\mathbb J}$ translates into four sets of equations
\begin{eqnarray}
&& C_{il}\nabla_l C_{jk}-B_{li}\partial_l C_{jk}+\circlearrowleft_{(ijk)}=0\quad\mbox{ for } i,j,k\in[1,M], \label{eq:JJ1}\\
&& C_{jl}\nabla_l B_{ki}-C_{il}\nabla_l B_{kj}+B_{li}\partial_l B_{kj}-B_{lj}\partial_l B_{ki}+B_{kl}\nabla_l C_{ij}\nonumber\\
&& \qquad \qquad \qquad +\bar{J}_{kl}\partial_l C_{ij}=0\quad\mbox{ for } i,j\in [1,M] \mbox{ and } k\in [1,N-M], \label{eq:JJ2}\\
&& B_{jl}\nabla_l B_{ki}-B_{kl}\nabla_l B_{ji}+\bar{J}_{jl}\partial_l B_{ki}-\bar{J}_{kl}\partial_l B_{ji}+C_{il}\nabla_l \bar{J}_{jk} \nonumber\\
&& \qquad \qquad \qquad -B_{li}\partial_l \bar{J}_{jk}=0 \quad \mbox{ for } i\in [1,M] \mbox{ and } j,k\in[1,N-M], \label{eq:JJ3}\\
&& B_{il}\nabla_l \bar{J}_{jk}+\bar{J}_{il}\partial_l \bar{J}_{jk} +\circlearrowleft_{(ijk)}=0\quad\mbox{ for } i,j,k\in[1,N-M], \label{eq:JJ4}
\end{eqnarray}
where $\circlearrowleft_{(ijk)} $ designates the terms obtained by circular permutations of the indices $(i,j,k)$, $\nabla_l=\partial/\partial \Phi_l$ and $\partial_l=\partial/\partial w_l$. So when the index $l$ is involved with $\nabla$, the implicit sum runs from $l=1$ to $M$. When it is involved with $\partial$, the implicit sum runs from $l=1$ to $N-M$.

Given Eqs.~(\ref{eq:Cas1})-(\ref{eq:Cas2}), the Jacobi identity for ${\mathbb J}_*$ reduces to 
\begin{equation}
( \bar{\mathbb J}+{\mathbb B}{\mathbb D}{\mathbb B}^\dagger)_{il}\partial_l ( \bar{\mathbb J}+{\mathbb B}{\mathbb D}{\mathbb B}^\dagger)_{jk}+ \circlearrowleft_{(ijk)}=0. 
\label{eq:JJstar}
\end{equation}
The aim is to use Eqs.~(\ref{eq:JJ1})-(\ref{eq:JJ4}) together with Eq.~(\ref{eq:Cas2}) in order to prove Eq.~(\ref{eq:JJstar}). 

Equation~(\ref{eq:JJstar}) can be decomposed into three sets of terms~:
\begin{eqnarray}
&& S_{ijk}=\bar{J}_{il}\partial_l \bar{J}_{jk}+({\mathbb B}{\mathbb D}{\mathbb B}^\dagger)_{il}\partial_l \bar{J}_{jk}+\circlearrowleft_{(ijk)} ,\\
&& T_{ijk}=\bar{J}_{il}\partial_l ({\mathbb B}{\mathbb D}{\mathbb B}^\dagger)_{jk}+\circlearrowleft_{(ijk)} ,\\
&& U_{ijk}= ({\mathbb B}{\mathbb D}{\mathbb B}^\dagger)_{il}\partial_l ({\mathbb B}{\mathbb D}{\mathbb B}^\dagger)_{jk}+\circlearrowleft_{(ijk)}.
\end{eqnarray}
Here we notice that all indices $i$, $j$, $k$ belong to $[1,N-M]$. 
Using Eq.~(\ref{eq:JJ4}), the $S$ terms can be rewritten as
$$
S_{ijk}=-B_{il}\nabla_l\bar{J}_{jk}+B_{im}D_{mn}B_{ln}\partial_l \bar{J}_{jk}+\circlearrowleft_{(ijk)}.
$$
By rewriting $B_{ln} \partial_l \bar{J}_{jk}$ using Eq.~(\ref{eq:JJ3}), a cancellation is obtained from  Eq.~(\ref{eq:Cas2}), and the $S$ terms are rewritten as
$$
S_{ijk}=B_{im}D_{im}(B_{jl}\nabla_l B_{kn}-B_{kl}\nabla_l B_{jn}+\bar{J}_{jl}\partial_l B_{kn}-\bar{J}_{kl}\partial_l B_{jn})+\circlearrowleft_{(ijk)}.
$$
By noticing a cancellation in the terms $BDJ\partial B$ in $S$ and $T$ (using a circular permutation of the indices $(i,j,k)$ and the antisymmetry of ${\mathbb D}$), we obtain
\begin{equation}
S_{ijk}+T_{ijk}=B_{im}D_{mn}(B_{jl}\nabla_l B_{kn}-B_{kl}\nabla_l B_{jn})+\bar{J}_{il}B_{jm}B_{kn}\partial_l D_{mn}+\circlearrowleft_{(ijk)}. \label{eq:ST}
\end{equation}

Concerning the $U$ terms, we decompose them into two parts~:
\begin{eqnarray*}
&& U^{(1)}_{ijk}=B_{im}D_{mn}B_{ln}D_{pq}B_{jp}\partial_l B_{kq} +B_{im}D_{mn}B_{ln}D_{pq}B_{kq}\partial_l B_{jp}+\circlearrowleft_{(ijk)},\\
&& U^{(2)}_{ijk}=B_{im}D_{mn}B_{ln}B_{jp}B_{kq}\partial_l D_{pq}+\circlearrowleft_{(ijk)}.
\end{eqnarray*}
The second term of $U^{(1)}$ is rewritten as $B_{jp}D_{pq}D_{mn}B_{im}B_{lq}\partial_l B_{kn}$ using a circular permutation of the indices $(i,j,k)$. Therefore, we have
$$
U^{(1)}_{ijk}=B_{im}D_{mn}D_{pq}B_{jp}(B_{ln}\partial_l B_{kq}-B_{lq}\partial_l B_{kn})+\circlearrowleft_{(ijk)}, 
$$ 
where we have used the antisymmetry of ${\mathbb D}$. From Eq.~(\ref{eq:JJ2}), $U^{(1)}$ is rewritten as
\begin{eqnarray}
U^{(1)}_{ijk}&=& -B_{im}D_{mn}({\mathbb B}{\mathbb D}{\mathbb C})_{jl} \nabla_l B_{kn}+B_{jm}D_{mn}({\mathbb B}{\mathbb D}{\mathbb C})_{il}\nabla_l B_{kn}\nonumber \\
&& -B_{im}B_{jp}B_{kl}D_{mn}D_{pq}\nabla_l C_{nq}-B_{im}B_{jp}D_{mn}D_{pq}\bar{J}_{kl}\partial_l C_{nq}+\circlearrowleft_{(ijk)}.\label{eq:U}
\end{eqnarray}
From Eq.~(\ref{eq:Cas2}) and using a circular permutation on the indices $(i,j,k)$, the first line of the previous equation cancels with the $BDB\nabla B$ terms in Eq.~(\ref{eq:ST}). Concerning the fourth term in Eq.~(\ref{eq:U}), we show that
$$
-\bar{J}_{kl}B_{jp}D_{pq} B_{im}D_{mn}\partial_l C_{nq}=-\bar{J}_{kl}B_{im}B_{jn}\partial_l D_{mn}.
$$
This property results from differentiating $B_{iq}-B_{im}D_{mn}C_{nq}=0$ with respect to $z_l$ together with Eq.~(\ref{eq:Cas2}) and the antisymmetry of ${\mathbb D}$. Using a circular permutation of $(i,j,k)$ these terms cancel with the $JBB\partial D$ terms of Eq.~(\ref{eq:ST}). 

After these steps, the Jacobi identity is rewritten as
\begin{equation}
\label{eq:STU}
S_{ijk}+T_{ijk}+U_{ijk}=B_{im}D_{mn}B_{ln}B_{jp}B_{kq}\partial_l D_{pq}-B_{im}B_{jp}B_{kl}D_{mn}D_{pq}\nabla_l C_{nq} + \circlearrowleft_{(ijk)}.
\end{equation}
Next the strategy is to get rid of the $\partial D$ terms. In order to do this, we insert $C$ terms through a $B$ coefficient in the first term of the previous equation, i.e., inserting $B_{jp}=B_{j\alpha}D_{\alpha \beta}C_{\beta p}$. From the identity
$$
C_{\beta p}\partial_l D_{pq} B_{kq}=\partial_l B_{k\beta} -\partial_l C_{\beta p} D_{pq} B_{kq}-C_{\beta p}D_{pq}\partial_l B_{kq},
$$ 
which is obtained by differentiating Eq.~(\ref{eq:Cas2}) with respect to $z_l$ (and using the antisymmetry of ${\mathbb D}$ and ${\mathbb C})$, the first term in Eq.~(\ref{eq:STU}) is rewritten as
\begin{eqnarray}
&& B_{im}D_{mn}B_{ln}B_{jp}B_{kq}\partial_l D_{pq}= -B_{im}D_{mn}B_{ln}B_{j\alpha}D_{\alpha\beta}D_{pq}B_{kq}\partial_l C_{\beta p}\nonumber \\
&&\qquad \qquad \qquad \qquad +B_{im}D_{mn}B_{ln}B_{j\alpha}D_{\alpha\beta}(\partial_l B_{k\beta}-C_{\beta p}D_{pq}\partial_l B_{kq}). \label{eq:last}
\end{eqnarray}
From Eq.~(\ref{eq:JJ1}) we replace $\partial C$ by $\nabla C$. The first term becomes
$$
B_{im}D_{mn}B_{j\alpha}D_{\alpha\beta}B_{kq}D_{qp} B_{ln}\partial_l C_{\beta p}+\circlearrowleft_{(ijk)}= B_{im}D_{mn}B_{j\alpha}D_{\alpha\beta}B_{kq}D_{qp} C_{nl}\nabla_l C_{\beta p} +\circlearrowleft_{(ijk)}.
$$
From the equation $B_{im}D_{mn}C_{nl}=B_{il}$ [see Eq.~(\ref{eq:Cas2})] we see that the previous term cancels with the second term in Eq.~(\ref{eq:STU}), still using a circular permutation of the indices $(i,j,k)$. In a similar way, the second term in Eq.~(\ref{eq:last}) vanishes by inserting $ B_{j\alpha}D_ {\alpha\beta} C_{\beta p}=B_{jp}$. Consequently, we have proved the Jacobi identity for the bracket~(\ref{eq:DB}) with Eqs.~(\ref{eq:Cas1})-(\ref{eq:Cas2}), i.e.,
$$
S_{ijk}+T_{ijk}+U_{ijk}=0. 
$$ 

\section{Examples}
\label{sec:examples}
\subsection{Example 1: Odd number of constraints}

First we describe a rather trivial example in order to illustrate the method. We consider the following Poisson matrix
$$
{\mathbb J}=\left( \begin{array}{ccccc} 0 & -z_3 & z_2 & 0 & 0\\ z_3 & 0 & -z_1 & 0 & 0\\ -z_2 & z_1 & 0 & 0 & 0\\ 0 & 0 & 0 & 0 & -1\\ 0 & 0 & 0 & 1 & 0
\end{array}\right),
$$
which corresponds to the Poisson bracket
$$
\{F,G\}=-{\bf z}\cdot \frac{\partial F}{\partial {\bf z}}\times \frac{\partial G}{\partial {\bf z}}-\frac{\partial F}{\partial w_1}\frac{\partial G}{\partial w_2}+\frac{\partial F}{\partial w_2}\frac{\partial G}{\partial w_1}.
$$
We impose three constraints $\Phi_k({\bf z},{\bf w})=z_k$ for $k=1,2,3$. The associated operator $\hat{\cal Q}$ is given by
$$
\hat{\cal Q}=\left( \begin{array}{ccccc} 1 & 0 & 0 & 0 & 0\\
0 & 1 & 0 & 0 & 0\\ 0 & 0 & 1 & 0 & 0
\end{array}\right).
$$
The matrix ${\mathbb C}$ is a $3\times 3$ antisymmetric, so it is not invertible. Its expression is
$$
{\mathbb C}=\left( \begin{array}{ccc} 0 & -z_3 & z_2 \\ z_3 & 0 & -z_1\\ -z_2 & z_1 & 0 \end{array} \right). 
$$ 
The following matrix ${\mathbb D}$ satisfies Eq.~(\ref{eq:forD})~:
$$
{\mathbb D}=\left( \begin{array}{ccc} 0 & 0 & 0\\ 0 & 0 & 1\\ 0 & -1 & 0
\end{array}\right).
$$
The solution of Eq.~(\ref{eq:forD}) is not unique, but all solutions lead to the same Dirac bracket with Poisson matrix given by Eq.~(\ref{eq:JJstar})~:
$$
{\mathbb J}_*=\left( \begin{array}{ccccc} 0 & 0 & 0 & 0 & 0\\
0 & 0 & 0 & 0 & 0\\ 0 & 0 & 0 & 0 & 0\\
0 & 0 & 0 & 0 & -1 \\ 0 & 0 & 0 & 1 & 0
\end{array}\right),
$$
which corresponds to the Dirac bracket
$$
\{F,G\}_*=-\frac{\partial F}{\partial w_1}\frac{\partial G}{\partial w_2}+\frac{\partial F}{\partial w_2}\frac{\partial G}{\partial w_1}.
$$

\subsection{Example 2: Vlasov-Poisson equation}

The second example concerns the Vlasov-Poisson reduction from the Vlasov-Maxwell equations. The field variables ${\bm\chi}({\bf z})=(f({\bf x},{\bf v}),{\bf E}({\bf x}),{\bf B}({\bf x}))$ where ${\bf z}=({\bf x},{\bf v})$, and the equations of motion are
\begin{eqnarray*}
&& \dot{f}=-{\bf v}\cdot \nabla f -({\bf E}+{\bf v}\times {\bf B})\cdot \partial_{\bf v} f,\\
&& {\dot {\bf E}}=\nabla \times {\bf B}-{\bf J},\\
&& \dot{\bf B}=-\nabla \times {\bf E},
\end{eqnarray*}
where ${\bf J}=\int d^3v \,{\bf v} f$. The Poisson bracket is given by
$$
\{F,G\}=\int d^6z F_{\bm\chi}\cdot {\mathbb J} G_{\bm\chi},
$$
where $F_{\bm\chi}$ is the functional derivative of the functional $F$ with respect to the field variables ${\bm\chi}$. The Poisson matrix is given by (see Refs.~\cite{morr80b,mars82,chan13})
$$
{\mathbb J}=\left( \begin{array}{ccc} -[f,\cdot] & -\partial_{\bf v} f & 0\\
-f\partial_{\bf v} & 0 & \delta({\bf v})\nabla \times \\
0 & -\delta({\bf v}) \nabla \times  & 0
\end{array}\right),
$$
where the small bracket $[\cdot,\cdot]$ is given by 
$$
[f,g]=\nabla f\cdot \partial_{\bf v}g-\partial_{\bf v}f\cdot \nabla g+{\bf B}\cdot(\partial_{\bf v}f\times\partial_{\bf v}g),
$$
where $\nabla$ (resp.~$\partial_{\bf v}$) is the partial derivative operator with respect to ${\bf x}$ (resp.~$\bf v$). 

In order to obtain the Vlasov-Poisson equations from the Vlasov-Maxwell equations we impose the following constraints:
\begin{equation}
\label{eq:constVM}
{\cal Q}[f,{\bf E},{\bf B}]({\bf x})=(\nabla \times {\bf E}, {\bf B}-{\bf B}_0({\bf x})),
\end{equation}
where ${\bf B}_0$ is a non-uniform background magnetic field. The operators $\hat{\cal Q}$ is given by
$$
\hat{\cal Q}=\left( \begin{array}{ccc} 0 & \nabla \times & 0\\ 0 & 0 & 1
\end{array}\right).
$$
The operator ${\mathbb C}$ is given by 
\begin{equation}
\label{eq:CVM}
{\mathbb C}=\left( \begin{array}{cc} 0 & (\nabla \times)^2\\ -(\nabla\times)^2 & 0
\end{array}\right). 
\end{equation}
The operator ${\mathbb C}$ is not invertible; however, the Dirac procedure still applies with an appropriate choice for ${\mathbb D}$ given by 
\begin{equation}
\label{eq:DVM}
{\mathbb D}=\left( \begin{array}{ccc} 0 & \Delta^{-1}\\
 -\Delta^{-1} & 0
\end{array}\right),
\end{equation}
so that Eq.~(\ref{eq:forD}) is satisfied. We notice that $\hat{\cal Q}({\mathbb 1}-{\mathbb D}{\mathbb C})\not= 0$ but ${\mathbb J}\hat{\cal Q}({\mathbb 1}-{\mathbb D}{\mathbb C})= 0$. This is due to the fact that $\nabla\cdot{\bf B}$ is a Casimir invariant of ${\mathbb J}$. 

As a result, the Poisson operator of the Vlasov-Poisson equations is given by 
$$
{\mathbb J}_*=\left( \begin{array}{ccc} 
-[f,\cdot] & -\nabla \Delta^{-1} \nabla \cdot \partial_{\bf v}f & 0\\
-\nabla \Delta^{-1} \nabla \cdot(f \partial_{\bf v}) & 0 & 0\\
0 & 0 & 0
\end{array}\right).
$$
It leads to the expression of the Poisson bracket~\cite{chan13}, 
\begin{eqnarray*}
\{F,G\}_*= \int d^6z \,f[F_f-\Delta^{-1}\nabla\cdot F_{\bf E},G_f-\Delta^{-1}\nabla\cdot G_{\bf E}].
\end{eqnarray*}
In the case of the constraints given by Eq.~(\ref{eq:constVM}), one of the constraints, $\nabla\cdot {\bf B}$, is already a Casimir invariant of the Vlasov-Maxwell bracket, and hence a first-class constraint. This kind of redundancy is a source for the non-invertibility of the matrix ${\mathbb C}$. In the present example, we eliminate this redundancy by modifying the set of constraints~(\ref{eq:constVM}), removing the divergence of ${\bf B}$, i.e., we consider the following constraints~:
$$
{\cal Q}[f,{\bf E},{\bf B}]({\bf x})=(\nabla \times {\bf E}, {\cal P}({\bf B}-{\bf B}_0({\bf x}))),
$$
where ${\cal P}$ is the projector on the solenoidal part given by ${\cal P}=1-\nabla \Delta^{-1}\nabla\cdot$. Using similar calculations as above, we find
$$
\hat{\cal Q}=\left( \begin{array}{ccc} 0 & \nabla \times & 0\\ 0 & 0 & {\cal P}
\end{array}\right),
$$
and ${\mathbb C}$ remains unchanged and is given by Eq.~(\ref{eq:CVM}). The matrix ${\mathbb D}$ given by Eq.~(\ref{eq:DVM}) satisfies $\hat{\cal Q}^\dagger (1-{\mathbb D}{\mathbb C})=0$. We have eliminated some redundancy at the origin of the non-invertibility of ${\mathbb C}$. As we shall see below, this kind of redundancy is not an issue in the proposed approach to compute the Dirac bracket. 

\section{Concluding remarks}
 
We consider an ensemble of $M$ constraints $\Phi_n$ of which the first $K$ ones are Casimir invariants of the original Poisson bracket (a particular family of first-class constraints). The matrix ${\mathbb C}$ is written as
$$
{\mathbb C}=\left( \begin{array}{cc} 0 & 0\\ 0 & \tilde{\mathbb C}
\end{array}\right),
$$
where $\tilde{\mathbb C}$ is assumed to be invertible. Given that $\Phi_k$ for $k=1,\ldots, K$ are Casimir invariants, ${\mathbb J}\hat{\cal Q}^\dagger = (0, {\mathbb J}\hat{\cal Q}_2^\dagger)$ where $\hat{\cal Q}_2$ is the matrix of the derivatives of $\Phi_{K+1},\ldots, \Phi_n$ with respect to the phase space variables. It is straightforward to show that any matrix of the form
$$
{\mathbb D}=\left( \begin{array}{cc} D_{11} & D_{12} \\ -D_{12}^\dagger & \tilde{\mathbb C}^{-1}
\end{array}\right),
$$
satisfies Eq.~(\ref{eq:forD}). Therefore in the definition of the constraints there is no need to worry about possible combinations giving rise to Casimir invariants of the original Poisson bracket. Notice that if some of the constraints are first-class but not Casimir invariants, this might lead to some inconsistencies in Eq.~(\ref{eq:forD}) like the ones in Refs.~\cite{dres88,Bhenn92}.

Another case of redundancy is when one (or several) constraints can be obtained from the other constraints. 
For instance, we assume that one constraint $\Phi_1$ is dependent of the other constraints, i.e., $\Phi_1=f(\Phi_2,\ldots,\Phi_M)$.
In this case, the usual Dirac's procedure cannot be carried out since the matrix ${\mathbb C}$ is not invertible. This matrix is given by
$$
{\mathbb C}=\left(\begin{array}{cc}
0 & -{\bm b}^\dagger\\ {\bm b} & \tilde{\mathbb C} 
\end{array}\right),
$$
where ${\bm b}=(\{\Phi_2,\Phi_1\},\{\Phi_3,\Phi_1\},\ldots,\{\Phi_M,\Phi_1\})^\dagger$ and the coefficients of $\tilde{\mathbb C}$ are $\{\Phi_n,\Phi_m\}$ for $n,m\geq 2$. In fact the vector $(-1,d_2,\ldots,d_M)$ where $d_m=\partial f/\partial \Phi_m$, belongs to the kernel of ${\mathbb C}$. We assume that $\tilde{\mathbb C}$ in invertible. Then the following matrix ${\mathbb D}$ satisfies Eq.~(\ref{eq:forD})
$$
{\mathbb D}=\left(\begin{array}{cc}
0 & -{\bm d}^\dagger\\ {\bm d} & \tilde{\mathbb C}^{-1}
\end{array}\right),
$$
where ${\bm d}=(d_2,\ldots,d_M)$. Here we have used the two properties, ${\bm b}^\dagger {\bm d}=0$ and $\tilde{\mathbb C}{\bm d}={\bm b}$. As a consequence this kind of redundancy is properly handled using Eq.~(\ref{eq:forD}) instead of the too stringent requirement that the matrix ${\mathbb C}$ is invertible.  

In summary the requirement for the existence of a Dirac-like bracket~(\ref{eq:DB}) is obtained by imposing Eq.~(\ref{eq:forD}), i.e., that the constraints are Casimir invariants. It translates into
\begin{equation}
\label{eq:condKer}
\mbox{Ker } {\mathbb C} \subset \mbox{Ker } {\mathbb J}\hat{\cal Q}^\dagger,
\end{equation}
which is a necessary and sufficient condition for the existence of a solution to Eq.~(\ref{eq:forD}). Indeed we denote $r$ the rank of ${\mathbb C}$ and we rewrite it as
$$
{\mathbb C}= O\left(\begin{array}{cc}
0 & 0\\ 0 & \tilde{\mathbb C}
\end{array}\right)O^\dagger,
$$
where $O$ is an orthogonal matrix and $\tilde{\mathbb C}$ is invertible. A possible antisymmetric solution to Eq.~(\ref{eq:forD}) is given by
$$
{\mathbb D}= O\left(\begin{array}{cc}
0 & 0\\ 0 & \tilde{\mathbb C}^{-1}
\end{array}\right)O^\dagger,
$$
given the condition~(\ref{eq:condKer}).

If the specific choice of constraints is such that Eq.~(\ref{eq:condKer}) is satisfied, then the Dirac bracket~(\ref{eq:DB}) can be computed from Eq.~(\ref{eq:forD}), and it is a Poisson bracket everywhere in phase space, and not just on the surface defined by the constraints. Otherwise some obstructions are present, and one should modify the set of constraints so as to reduce the kernel of ${\mathbb C}$.  

\section*{Acknowledgments}
This work was supported by the Agence Nationale de la Recherche (ANR GYPSI) and by the European Community under the contract of Association between EURATOM, CEA, and the French Research Federation for fusion study. The views and opinions expressed herein do not necessarily reflect those of the European Commission. CC acknowledges fruitful discussions with P.J. Morrison and with the \'Equipe de Dynamique Nonlin\'eaire of the Centre de Physique Th\'eorique of Marseille.


\section*{References}

\begin{thebibliography}{10}

\bibitem{dira50} P.A.M. Dirac, Can. J. Math. 2 (1950) 129. 

\bibitem{dira58} P.A.M. Dirac, Proc. Roy. Soc. Lond. A 246 (1958) 326.

\bibitem{Bhans76} A. Hanson, T. Regge, C. Teitelboim, Constrained Hamiltonian Systems, Accademia Nazionale dei Lincei, Roma, 1976.

\bibitem{Bsund82} K. Sundermeyer, Constrained Dynamics, Springer-Verlag, Berlin, 1982.

\bibitem{morr09} P.J. Morrison, N. Lebovitz, J. Biello, Ann. Phys. 324 (2009) 1747.

\bibitem{Bhenn92} M. Henneaux, C. Teitelboim, Quantization of Gauge Systems, Princeton University Press, Princeton, New Jersey, 1992. 

\bibitem{abre02} E.M.C. Abreu, D. Dalmazi, E.A. Silva, Int. J. Mod. Phys. A 17 (2002) 395. 

\bibitem{bizd07} C. Bizdadea, E.M. Cioroianu, S.O. Saliu, S.C. Sararu, O. Balus, J. Phys. A: Math. Theor. 40 (2007) 14537. 

\bibitem{deri96} A.A. Deriglazov, A.V. Galajinsky, S.L. Lyakhovitch, Nucl. Phys. B 473 (1996) 245.

\bibitem{Bmars02} J.E. Marsden, T.R. Ratiu, Introduction to Mechanics and Symmetry, Springer-
Verlag, Berlin, 2002.

\bibitem{morr82} P.J. Morrison, in Mathematical Methods in Hydrodynamics and Integrability in Related Dynamical Systems, La Jolla Institute, 1981, edited by M. Tabor and Y.M. Treve, AIP Conf. Proc. 88 (1982) 13.

\bibitem{morr98} P.J. Morrison, Rev. Mod. Phys. 70 (1998) 467.

\bibitem{chan13} C. Chandre, L. de Guillebon, A. Back, E. Tassi, P.J. Morrison, J. Phys. A: Math. Theor. 46 (2013) 125203.

\bibitem{morr80b} P.J. Morrison, Phys. Lett. A  80 (1980) 383.

\bibitem{mars82} J.E. Marsden, A. Weinstein, Physica D 4 (1982) 394.

\bibitem{dres88} A. Dresse, J. Fisch, M. Henneaux, C. Schomblond, Phys. Lett. B 210 (1988) 141.

\end{thebibliography}
\end{document}